\documentclass[seceq]{ptptex}

\title{Extremal dilatonic black holes in 4D Gauss-Bonnet gravity}

\author{Chiang-Mei \textsc{Chen}~\footnote{Email:
cmchen@phy.ncu.edu.tw}}

\inst{Department of Physics and Center for Mathematics and
Theoretical Physics, \\ National Central University, Chungli 320,
Taiwan}

\abst{This is a report of our recent investigation\cite{Chen:2006ge}
on the extremal dilatonic black holes in four dimensional
Gauss-Bonnet gravity. We found that a global solution can exist only
when the dilaton coupling is less than a critical value which can be
determined numerically. Moreover, the black hole horizon is
stretched by the Gauss-Bonnet correction and the entropy is twice
the value given by Bekenstein-Hawking formula.}

\begin{document}

\maketitle

\section{Introduction}
It is well known that the spherical symmetric charged black holes of
the Einstein-Maxwell gravity are described by the
Reissner-Nordstr\"om (RN) solutions. This class of solutions carries
two conserved quantities, namely mass $M$ and electric charge $Q$,
and they are asymptotically flat. There are two essential geometric
characteristics of those spherical charged black holes. First, there
exist a singular point (singularity) at $r = 0$ where the curvature
diverges. Second, there are, in general, two apparent singular
2-surfaces located at $r_{\pm} = M \pm \sqrt{M^2 - Q^2}$. Indeed,
these two surfaces are coordinate singularities which correspond to
the outer (event) and inner (Cauchy) horizons. The existence of an
(event) horizon is physically crucial, inspiring the cosmic
censorship conjecture: the singularity should be protected by
(event) horizon.

The black hole mechanism behaves like a thermal system. The
connection was first observed by Bekenstein via comparing the area
increasing theorem of black holes and the 2nd law for entropy in
thermodynamics. Soon the other corresponding thermodynamical laws
also were derived from black hole mechanisms. The corresponding
macroscopic thermal quantities for a black hole, such as the Hawking
temperature $T$ and entropy $S$, are given by
\begin{equation}\label{TS}
T = \frac{\kappa}{2\pi}, \qquad S = \frac{A}4,
\end{equation}
where $\kappa$ is the surface gravity on the horizon and $A$ is the
area of horizon. The non-vanishing temperature indicates that the
black hole is unstable and should emit thermal radiation (due to the
quantum effect).

Generically, charged black holes can have two horizons: inner and
outer. For the extremal limit ($M^2 = Q^2, r_+ = r_- = M$), the two
horizons degenerate and the Hawking temperature vanishes ($\kappa =
(r_+ - r_-)/2 r_+^2$). However, the entropy (area of horizon) is
non-vanishing ($r_H = M$) which represents the quantum degrees of
freedom inside the black hole.

\section{Dilatonic charged black holes}
There are three possible generalizations of black hole solutions by
introducing new ingredients: (i) dilaton fields, for example
Brans-Dicke theory, (ii) extra dimensions, such as Kaluza-Klein
theory, (iii) higher-rank form fields, extending black holes to
black branes. Those new ingredients are all essential in low energy
effective string theory. Moreover, the extremal black holes
correspond to some BPS configurations which generically preserve
partial supersymmetry. Here we will focus on extremal black holes
with a dilaton field extension.

The action of the four dimensional Einstein-Maxwell-dilaton gravity
is
\begin{equation}\label{action}
S = \frac1{16\pi} \int d^4 x \sqrt{-g} \left( R - 2 (\partial
\phi)^2 - {\rm e}^{ 2 a \phi} \, F_{[2]}^2 \right).
\end{equation}
The metric of the spherical symmetric dilatonic black holes for the
particular value of dilaton coupling $a = -1$ is
\begin{equation}
ds^2 = - f(r) dt^2 + \frac{dr^2}{f(r)} + R^2(r) \, d\Omega_2^2,
\qquad f(r) = 1 - \frac{r_+}{r}, \quad R^2(r) = r (r - r_-),
\end{equation}
and the mass and charge are related to the radii of the inner and
outer horizons by $M = r_+/2, Q^2 = r_+ r_- / 2$. Here we leave out
the explicit expressions for the dilaton and gauge fields, which are
not essential in our later discussion. The complete solution and
also the general solutions with arbitrary $a$ can be found in the
references\cite{Gibbons:1987ps}.

From the function $R(r)$ one can easy realize that, in the extremal
limit, i.e. $r_+ = r_-$, the degenerated event horizon shrinks to a
point, $R(r_+ = r_-)= 0$, which implies that the entropy of black
hole vanishes. The result of zero entropy raises a puzzle that the
expected quantum degrees of freedom inside the back holes seem to
disappear completely. This discrepancy comes from the fact that
general relativity is not sufficient to catch the fundamental point
in this case, and higher curvature corrections are necessary to
stretch the horizon and reproduce the correct entropy corresponding
to the microstate degrees of freedom. We investigate this issue by
considering the correction coming from the Gauss-Bonnet combination
of quadratic curvature \cite{Chen:2006ge}.

The action of four dimensional Gauss-Bonnet gravity can be obtained
from (\ref{action}) simply by replacing $F_{[2]}^2$ with $F_{[2]}^2
- {\alpha {\cal L}_{\rm GB}}$ where the Gauss-Bonnet term is defined
by ${\cal L}_{\rm GB} = R^2 - 4 R_{\mu\nu} R^{\mu\nu} +
R_{\alpha\beta\mu\nu} R^{\alpha\beta\mu\nu}$ and $\alpha$ is its
coupling. The ansatz for the metric and Maxwell field potential
($F_{[2]} = d A_{[1]}$) are
\begin{equation}
ds^2 = - w(r) dt^2 + \frac{dr^2}{w(r)} + \rho^2(r) d\Omega_2^2,
\qquad A_{[1]} = - f(r) \, dt - q_m \cos\theta \, d\varphi.
\end{equation}
Here $q_m$ is the magnetic charge parameter, and the electric part
can be directly solved and then the electric charge parameter $q_e$
is introduced as
\begin{equation}
f'(r) = q_e \rho^{-2} {\rm e}^{- 2 a \phi}.
\end{equation}

In Gauss-Bonnet gravity, the field equations remain second order in
the metric, linear in the second derivative. However, the equations
of motion become rather complicated. Our approach is to consider the
analytical expansions near the horizon and at infinity, and then
numerically interpolate these two sets of boundary date. So, lets
first consider the series expansions around a particular value $r =
r_H$ (supposed to be a horizon) in powers of $x = r - r_H$ (here
$P(r) := {\rm e}^{2 a \phi(r)}$) as
\begin{equation}
w(r) = \sum_{{k=2}}^\infty w_k x^k, \qquad \rho(r) =
\sum_{k=0}^\infty \rho_k x^k, \qquad P(r) = \sum_{k=0}^\infty P_k
x^k.
\end{equation}
The function $w$ starts from the quadratic term (vanishing of $w_0$
means that $r=r_H$ is a horizon, vanishing of $w_1$ means that the
horizon is degenerate). The electric charge is then given by
\begin{equation}
q_e = \frac{\sqrt{4 \alpha + q_m^2}}{2 \alpha + q_m^2} \;
\frac{\rho^2_0}2.
\end{equation}
Any solution with finite horizon radius $\rho_0$ must have non-zero
electric charge, depending on the magnetic charge. With fixed
$\rho_0$, $q_e$ decreases with increasing $q_m$ and approaches zero
in the limit $q_m\to\infty$.

For simplicity, we only focus on purely electric charged black
holes. The near horizon expansion for the electric solution is
\begin{eqnarray}
w(r) &\simeq& \frac1{\rho_0^2} \left[ x^2 - \frac{2 (5 a^2 - 3)}{3}
\left( \frac{\alpha P_1}{ a^2\rho_0^2} \right) x^3 \right] + O(x^4),
\\
\rho(r) &\simeq& {\rho_0} \left[ 1 +  ({a^2 - 1}) \left(
\frac{\alpha P_1}{{a^2}\rho_0^2} \right) x - \frac{2 a^2 (a^4 - 6)}{
(5 a^2 - 3)} \left( \frac{\alpha P_1}{{a^2}\rho_0^2} \right)^2 x^2
\right] + O(x^3),
\\
P(r) &\simeq& \frac{\rho_0^2}{\alpha} \left[ \frac14  + a^2 \left(
\frac{\alpha {P_1}}{a^2 \rho_0^2} \right) x + \frac{a^2(a^4 - 5 a^2
- 3)}{ (5a^2 - 3)} \left( \frac{\alpha P_1}{ a^2 \rho_0^2} \right)^2
x^2 \right] + O(x^3).
\end{eqnarray}
There are two independent parameters, $\rho_0, P_1$, in the
expansion and the near horizon geometry is $AdS_2 \times S^2$. The
asymptotic expansion, for asymptotically flat, is
\begin{eqnarray}
w(r) &=& 1 - \frac{2M}r + \frac{\alpha Q_e^2}{r^2} + O(r^{-3}),
\\
\rho(r) &=& r - \frac{D^2}{2 r} - \frac{D(2 M D - \alpha a Q_e^2)}{3
r^2} + O(r^{-3}),
\\
\phi(r) &=& \phi_{\infty} + \frac{D}r + \frac{2 D M - \alpha  a
Q_e^2}{2 r^2}  + O(r^{-3}),
\end{eqnarray}
where $M, D, \phi_\infty$ are the mass, dilaton charge, asymptotic
value of dilaton, and the electric charge is given by $Q_e = q_e
\mathrm{e}^{-a\phi_\infty}$.

Our numerical investigation \cite{Chen:2006ge} shows that the value
of $P_1$ is fixed (depending on $a, \alpha$) in order to get an
asymptotically flat solution. The physical quantities mass $M$,
dilaton charge $D$, and asymptotic value of dilaton $\phi_\infty$
are all determined only by the value of the parameter $\rho_0$ (i.e.
charge). Moreover, we also found that the global solution can exist
only when the dilaton coupling $a$ is less than a critical value
$a_{\rm cr} \simeq 0.488219703$. If $a > a_{\rm cr}$ a singularity
appears outside horizon. The physics behind the existence of such a
critical value is unclear.

The Bekenstein-Hawking entropy-area formula in (\ref{TS}) breaks
down for the theories including higher curvature terms. Formally,
the general expression of entropy can be obtained via Wald's Noether
charge approach \cite{Wald:1993nt}. However, for the black holes
with a near horizon geometry of $AdS_2 \times S^{D-2}$, the
calculation can be significantly simplified and the entropy can be
obtained by the near horizon data via the entropy function proposed
by Sen \cite{Sen:2004dp}. Following Sen's approach, the entropy of
an extremal dilatonic Gauss-Bonnet black hole is $S = 2 \pi \rho_0^2
= A/2$. Unexpectedly, the higher curvature terms contribute an equal
amount of entropy as the Hilbert-Einstein action (scalar curvature).

\section{Discussion}
By considering the extremal dilatonic black hole, we found the
higher curvature corrections are required. A particular interesting
case is to include the Gauss-Bonnet term in four dimensional dilaton
gravity. In this case, the black hole solutions form a one-parameter
family and exist in a finite range of the dilaton coupling constant
$a$. The corresponding extremal solution in the
Einstein-Maxwell-dilaton theory (\ref{action}) has two parameters
($q_e$ and $\phi_\infty$). Thus, the asymptotic value of the dilaton
is no longer a free parameter when the Gauss-Bonnet term is
included.

The entropy calculated by Sen's entropy function approach is twice
the value from the Bekenstein-Hawking formula. This new entropy-area
relation has been checked in theories with more general higher
curvature corrections \cite{CCMOP}. The existence of the threshold
value of the dilaton coupling constant under which the global
solutions cease to exist is an interesting new phenomenon which may
be related to the string-black hole transition. We think that our
model can be regarded as simple toy model describing the
string-black hole transition.

\section*{Acknowledgements}
The result reported here is based on a collaboration
\cite{Chen:2006ge} with D.~V.~Gal'tsov and D.~G.~Orlov. The author
would like to thanks the organizers of ICGA8 for the invitation to
attend this wonderful conference. This work was supported by the
National Science Council under the grant NSC 96-2112-M-008-006-MY3,
and in part by the National Center for Theoretical Sciences.

%

\end{document}